\documentclass[preprint]{aastex}

\received{2003 August 19}
\begin{document}
 
\title{THE PULSAR-WHITE DWARF-PLANET SYSTEM IN MESSIER 4: IMPROVED ASTROMETRY $^1$}

\author{
Harvey B. Richer\altaffilmark{2},
Rodrigo Ibata\altaffilmark{3},
Gregory G. Fahlman\altaffilmark{4},
Mark Huber\altaffilmark{2}
}

\slugcomment{\it Submitted to The Astrophysical Journal Letters August 19, 2003. 
 }

\lefthead{Richer {\it et al.}}
\righthead{Triple system in M4}

\altaffiltext{1}
{Based on observations with the NASA/ESA Hubble Space Telescope, obtained at
the Space Telescope Science Institute, which is operated by AURA under NASA
contract NAS 5-26555. Based on observations with the Canada-France Hawaii Telescope, which is operated by the National Research Council of Canada, the Centre National de la Recherche Scientifique of France and the University of Hawaii.}
\altaffiltext{2}
{Department of Physics \& Astronomy, University of British Columbia,
6224 Agricultural Road, Vancouver, BC V6T 1Z1, Canada. richer@astro.ubc.ca,
mhuber@astro.ubc.ca}
\altaffiltext{3}
{Observatoire de Strasbourg, 11 rue de l'Universite, F-67000 Strasbourg, France.
ibata@newb6.u-strasbg.fr}
\altaffiltext{4}
{National Research Council, Herzberg Institute of Astrophysics, 5071 West
Saanich Road, RR5, Victoria, BC V9E 2E7, Canada. greg.fahlman@nrc-cnrc.gc.ca}

 \vspace{.2in}

\begin{abstract}
A  young and  undermassive  white  dwarf has  been  identified as  the
possible companion to the  millisecond pulsar PSR B1620-26 in Messier
4. This association is important as it then helps constrain
the mass of the third
body in  the system to  be of order  a few times that  of Jupiter.
The presence of this planet in M4 has critical implications
for planetary formation
mechanisms in metal-poor environments such as globular clusters and the 
early Universe. The  identification of the  white dwarf is purely  via 
the
agreement in position between it and  the pulsar and was limited by the
accuracy of  the pointing of HST  which is $ \pm  0.7\arcsec$. We have
redetermined the position  of the white dwarf using  ground-based data
tied to  USNOB-1.0 and find  that the pulsar  and white dwarf  are now
coincident to within $0.12 \pm 0.13\arcsec$ further strengthening the
case  for association  between  the  two. We  have  also attempted  to
improve the proper motion measurement  of the white dwarf by a maximum
likelihood analysis of the  stellar positions measured over a baseline
of 5 years. While the errors are  reduced by almost a factor of 6 from
our previous work, we still  have not resolved the cluster's intrinsic
dispersion  in proper  motion.  Thus  the proper  motion of  the white
dwarf with respect  to the cluster itself is  still not known although
it  is very  small and  is  within $2\sigma$  of that  of the  cluster
internal dispersion.
\end{abstract}


\section{Introduction}

Messier 4 contains a unique stellar/planetary system. It consists of a
central tight binary composed of  an 11 msec pulsar (PSR B1620-26) and
a stellar  companion (thought to be  the white dwarf that  spun up the
neutron  star) together  with  a distant  object  possessing either  a
planetary mass in a low eccentricity orbit or a stellar companion in a
highly eccentric one  (Lyne {\it et al.}\ 1987;  McKenna \& Lyne 1988;
Backer, Foster  \& Sallmen  1993, Michel 1994, Rasio 1994, Thorsett {\it  et al.}\  1999, Ford,
Joshi \&  Rasio 2000). The former  scenario was deemed early  on to be
the more probable one. The existence of this
triple system  is of great cosmogonic significance  as, if primordial,
it could  suggest the presence  of numerous solar systems  that formed
early in the history of the Universe.

The close-in  stellar companion to PSR B1620-26  had escaped detection
until recently  when Sigurdsson  {\it et al.}\  (2003) showed  that an
undermassive  and  young  white  dwarf  was the  likely  partner.  The
properties of  this object seemed  to fit a pre-existing  scenario for
the history of this triple wherein an exchange reaction in the core of
M4  captured  both the  white  dwarf's  progenitor  and its  planetary
companion into an orbit around  a neutron star (Sigurdsson 1992, 1993,
1995).

While  the identification  of the  companion to  the pulsar  is highly
suggestive, it does have  its limitations. The positional agreement is
constrained by  how well  the coordinates of  the HST guide  stars are
known. The  uncertainty in their  location limits the accuracy  of the
absolute position of  the white dwarf to $\pm  0.7\arcsec$. The pulsar
position by  contrast is known to about  $0.003\arcsec$ through pulsar
timing (Thorsett {\it et al.} 1999).  The system should have  been 
ejected on almost a radial orbit
from the  core after the  exchange interaction.  The  energy available
for  this recoil  originated in  the increased  binding energy  of the
newly formed binary as the new stellar companion was more massive than
the original one.  The recoil velocity of the system is expected to be
$10 - 15$  km/sec (Sigurdsson 1993) which is  near the escape velocity
from the cluster.   The orbit of the triple  system within the cluster
evolved  via  dynamical  friction  and  orbital diffusion  and  it  is
expected to return  to the cluster core on a  timescale of $\sim 10^9$
yrs (Sigurdsson  {\it et  al.}\ 2003). This  suggests that  the system
could have a somewhat higher velocity with respect to the cluster than
equilibrium dynamics would suggest. It is of interest then to look for
this in the system's proper motion.

Both the  position and  proper motion of  the system are  addressed in
this {\it Letter} with  improved measurements of both. The significant
orbital velocity of the white dwarf around the neutron star could also
be used  as a definitive test  of the association between  the two but
this awaits future observations.

\section{The Data}

The  data   analyzed  are  composed  of  both   ground-based  and  HST
images. The ground-based observations are short exposure 
(5$\times2$ sec averaged into a master image with high pixels removed
by an average sigma clip) CFHT
12K I-band  data centered on M4. The complete HST  data set analyzed
here consists solely  of images in F814W so that effects of different
bandpasses  play  no  part  in  the  analysis.  The  earliest  images,
9$\times$800 sec  in F814W, were  secured in 1995  as part of  GO 5461
from cycle  4 (Richer {\it  et al.}\ 1995,  1997; Ibata {\it  et al.}\
1999). The second epoch exposures came from the HST archives (GO 8153)
and   consist  of   8$\times$700  sec   exposures  giving   images  of
approximately  equal depth  to  the  cycle 4  ones.   All images  were
secured with  the same  roll angle and  in all cases  the pulsar/white
dwarf fall on  WF3. The images from GO 6166 obtained  in 1999 were not
used as they  were much shallower than those  mentioned above and were
secured  so that the  pulsar/white dwarf  were on  the PC.  This could
cause  further difficulties  with potential  proper  motion reductions
from WF data because of the  scale difference. The HST images were all
preprocessed according  to the recipe  given in Stetson {\it  et al.}\
1998.

\section{Positional Astrometry}

In order to  construct an appropriate reference frame,  we carried out
photometry on the  CFHT images using the DAOPHOT  (Stetson 1987) suite
of programs. In total 16240 stars in the direction of M4 were measured
on a  single CFHT 12K chip.  The  location of these stars  is shown in
Figure 1.  These were  converted into standard coordinates centered on
$16^h23^m35.41^s$ $-26^{\circ}31\arcmin31.9\arcsec$ (J2000). 
Within this area of the
chip we  also selected all of  the stars on the  USNOB-1.0. There were
816 of  these (circled in  Figure 1).  We then  cross-identified these
two lists, keeping only those  stars that had no neighbors closer than
$3.0\arcsec$,  that  had $I  <  18$  and  whose photometry  satisfield
$|I_{CFHT}  - I_{USNO}|  <  1.5$. This  latter  constraint may  appear
overly generous,  but the USNO photometry  near the faint  end of this
range  is  rather  poor.    We  iteratively  refined  a  6-coefficient
polynomial  transformation while rejecting  cross-identifications with
large residuals.  This resulted in 274 cross-identified stars, with an
RMS in the ensuing transformations of $0.12\arcsec$.

This  process  was then  repeated  using the  CFHT  data  (now on  the
USNOB-1.0 system) and  our existing HST WF3 astrometry  (Ibata {\it et
al.}\ 1999). In this case any  object detected on both the CFHT images
and the HST ones were  used without regard to their relative magnitude
measurements. Here  428 stars were  matched (larger dots in  Figure 1)
and  the  resultant  RMS  in  the  transformation  was  $0.06\arcsec$.
Combining the 2 transformations gave a position for the white dwarf at
$16^h23^m38.217^s$ $-26^{\circ}31\arcmin53.662\arcsec$,  that is $0.126 \pm0.13\arcsec$
from  the  pulsar.   This  result  does  not  include  any  systematic
uncertainty coming from the USNOB-1.0.

Note from these results that the red main sequence star mentioned as a
possible pulsar companion  in Sigurdsson {\it et al.}\  (2003), is now
about  $6\sigma$ away  from the  nominal pulsar  position and  is very
unlikely to be associated with the pulsar/planetary system.

\section{Proper Motion Measurements}

The  velocity dispersion  near  the core  of  M4 is  about 3.5  km/sec
(Peterson, Rees  \& Cudworth 1995). At  the distance of  M4 (1.73 kpc)
this corresponds to a  (2-dimensional) proper motion dispersion of 0.6
mas/yr. Our measurements in the  inner (Richer {\it et al.}\ 2003) and
outer (Richer {\it et al.}\  2002) regions of the cluster have yielded
errors in  the range of  3 mas/yr for  faint stars. As was shown 
originally by 
Cudworth and Rees (1990), M4 has a high proper motion with respect
to the field population. Building on this, our goal  has been
simply to  be able to distinguish  cluster from field  stars so as to
produce proper motion selected color magnitude diagrams. However, Bedin
{\it  et al.}\  2003 are  interested in  the
internal dynamics of globular clusters and have a long term program to
resolve the dispersion. They currently claim  a measurement precision
of about 0.02 pixels which, when coupled with long baselines, may make
this goal eventually possible.

The measurement uncertainty which we have achieved in our earlier work
is about  0.15 pixels  for stars down  to $V  = 27.5$ (Richer  {\it et
  al.}\ 2003). With the more sophisticated analysis discussed below we
improve this to about 0.03 pixels at the expense of not reaching stars
as faint. However,  this is not a concern for  the present analysis as
the white dwarf has $V = 24.0$.

A reference frame  for the cluster was constructed  by first selecting
only cluster members from our earlier work. Photometry was carried out
for these objects on all  the individual HST frames using ALLSTAR. The
frames were  then transformed onto the  sky using the  Trauger {\it et
al.}\ (1995)  corrections.  This effectively corrects  for the optical
distortions in the WFPC2 camera.  All the frames from both epochs were
then  transformed  to a  single  reference  frame.   The RMS  of  this
solution was typically 0.08 pixels, and with generally about 500 stars
in each frame, this resulted  in a positional uncertainty of about 3.6
mas.

Using  the  maximum likelihood  method  described  in  Ibata \&  Lewis
(1998),  we  find the  positions  of the  white  dwarf  and red  dwarf
mentioned  in Sigurdsson  {\it  et  al.}\ (2003)  as  listed in  Table
1. From this Table we derive that the proper motion of the white dwarf
companion to  PSR B1620-26 with respect  to the cluster  is about $0.9
\pm 1.1$  mas/yr. Similarly  the red main  sequence star has  a proper
motion of  $1.1 \pm 1.0$  mas/yr so that neither  exhibits significant
peculiar  motion with respect  to M4.  In fact  if the  cluster proper
motion dispersion is 0.6 mas/yr  as expected then neither exceeds this
by more than about $2\sigma$.

\section{Other Considerations}

It  is clear  that a  much longer  baseline (about  15 years)  than we
currently have will  be required to both resolve  the M4 proper motion
dispersion and measure the motion  of the white dwarf companion to PSR
B1620-26  with respect  to  the  cluster. This  may  eventually be  an
important check  on the dynamical  history of this  system (Sigurdsson
{\it et al.}\ 2003, Sigurdsson 1992, 1993, 1995). However, other tests
are currently possible.  With an orbital period of  only 191 days, the
white dwarf  will exhibit  a velocity amplitude  of about  40 km/sec
which should, with a determined  effort, be capable of detection. This
velocity curve will be  critical  datum   toward  an  independent
measurement of the mass of its  Pop II neutron star companion. A spectroscopic mass determination of the white dwarf itself would also provide an important
confirmation of the mass estimated from fitting its location in the M4 CMD
to model cooling curves (Sigurdsson {\it et al.} 2003).

The case for the association
of the white dwarf with PSR B1620-26 has been strongly strengthened
by the present results, supporting the conclusion that the third body
is most likely
of planetary  mass (Sigurdsson {\it et al.} 2003). Since  this system has
the metal abundance of  M4 (30 times less than that of  the Sun) it is
strongly suggestive that  a mode of planetary formation that does not
rely on planetesimals  has been in operation as the entire proto-planetary
disk out of which this object formed would only have contained a few Earth
masses of metals. Such  models, relying on
disk  instability mechanisms,  have  been suggested  for  more than  a
decade  now and  were recently  reviewed  by Boss  (2002). This  would
produce planets much earlier in  the Universe than one which relied on
a mechanism that required  high metallicity. The implications here for
a potential early rise to life in the Universe are thus obvious.

\begin{acknowledgements}
The  research of  HBR and  GGF  is supported  in part  by the  Natural
Sciences and  Engineering Research Council of Canada.  HBR extends his
appreciation to the  Killam Foundation and the Canada  Council for the
award of a Canada Council Killam Fellowship.  \end{acknowledgements}

\clearpage

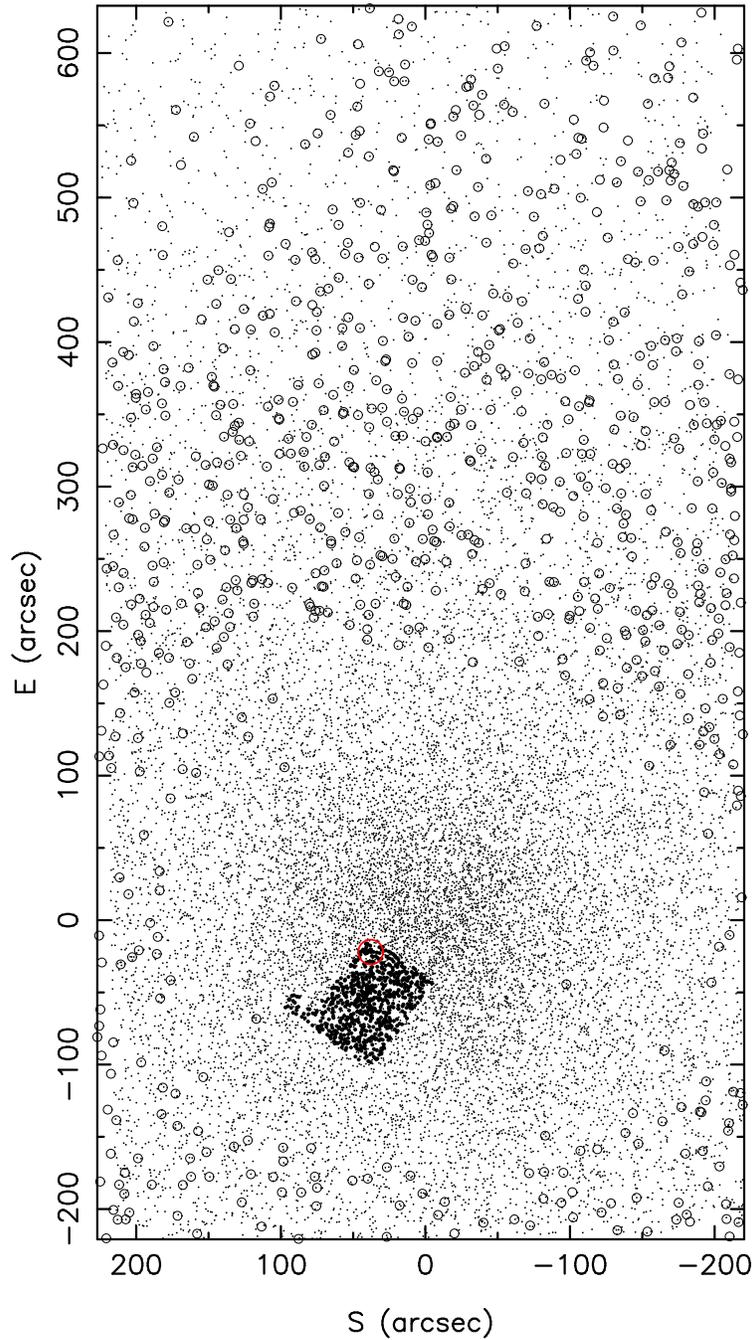
\figcaption[f1.eps]{All the  stars identified on  the CFHT
exposure (black  dots) and those  in common with  USNOB-1.0 (circled).
The inner darker  points locate the HST M4  field (WF3) containing PSR
B1620-26.  The red circle is centered on the pulsar position and has a
radius  of $10\arcsec$.   The somewhat  odd shape  of the  WF3 sources
displayed here  comes from  the stars being  selected solely  from the
inner M4 field of Ibata {\it et  al.} (1999) with the overlap stars from the
more distant field not being included. \label{cfht}}

\epsscale{0.6}
\plotone{f1.eps}

\clearpage

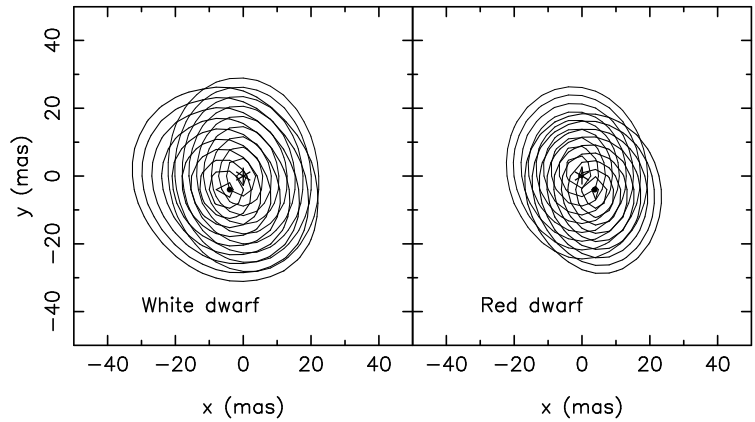
\figcaption[f2.eps]{The  maximum likehood contours  for the
positions of  the white dwarf and  red star. The positions are relative
to the centroid for the 1995 data. The absolute astrometry is significantly
less precise and has an uncertainty of 130 mas. 
Contours from the  1995 data are displayed with  thin lines, while the
thick lined countours correspond to  the 2000 data. The asterisk marks
the most likely position of  the point-source centroid given the first
epoch data, while the heavy dot  shows the same point given the second
epoch data.   The contour  intervals are such  that the  $n$th contour
marks the boundary of the region  where the likelihood has fallen by a
factor of $\exp{-{{n^2} \over 2}}$  from the most likely value (so the
image centroid is $< 10^{-22}$ times less likely to be situated beyond
the tenth contour than at the contour peaks).
\label{like}}
\plotone{f2.eps}

\clearpage

\begin{deluxetable}{ccccc}
\tablecolumns{5}
\tablewidth{0pc}
\tablecaption{Stellar Positions and Uncertainties \label{table1}}
\tablehead{
\colhead{Epoch} &  \colhead{x} & \colhead{y} &
 \colhead{dx} & \colhead{dy} \\
 }

\startdata
&  &White Dwarf &  &  \nl
1995 & 144.426 & 322.535 & 0.0228 & 0.0295 \nl
2000 & 144.386 & 322.510 & 0.0257 & 0.0301 \nl
& &Red Star &  & \nl
1995 & 138.207 & 325.304 & 0.0217 & 0.0288 \nl
2000 & 138.242 & 325.259 & 0.0197& 0.0243 \nl
\enddata
\end{deluxetable}
\end{document}